\pdfoutput=1
\documentclass[conference]{IEEEtran}
\usepackage{cite}
\usepackage[T1]{fontenc}
\usepackage[utf8]{inputenc}
\usepackage{microtype}
\usepackage{inconsolata}
\usepackage{graphicx}
\usepackage{amsmath}
\usepackage{booktabs}
\usepackage{tabularx}
\usepackage{tikz}
\usepackage[hyphens]{url}
\usepackage{float}
\makeatletter
\newcommand\floatc@IEEEtop[2]{{\centering\footnotesize #1\par}\vspace{1pt}{\centering\footnotesize\scshape #2\par}}
\newcommand\fs@IEEEtop{\def\@fs@cfont{\rmfamily}\let\@fs@capt\floatc@IEEEtop
  \def\@fs@pre{}\def\@fs@post{}\def\@fs@mid{\vspace{5pt}}\let\@fs@iftopcapt\iftrue}
\makeatother
\floatstyle{IEEEtop}\restylefloat{table}
\usepackage[hidelinks]{hyperref}
\hypersetup{
  pdftitle={Collective Reasoning, Single-Writer Execution: Fusion-MoA for Long-Horizon Agents},
  pdfauthor={Zongyou Yang, Yinghan Hou, Botao Su, Ercheng Zhao}}

\definecolor{FusionGold}{HTML}{B8891E}
\DeclareRobustCommand{\circled}[1]{\tikz[baseline=(c.base)]{\node[circle,fill=FusionGold,text=white,inner sep=0.4pt,minimum size=1.05em,font=\scriptsize\bfseries\sffamily](c){#1};}}
\newcommand{\Exec}{D12}
\newif\ifanon
\anonfalse
\newcommand{\citep}{\cite}
\newcommand{\citealp}{\cite}
\let\appendix\appendices
\title{Collective Reasoning, Single-Writer Execution:\\Fusion-MoA for Long-Horizon Agents}

\author{\IEEEauthorblockN{Zongyou Yang\IEEEauthorrefmark{1}\IEEEauthorrefmark{3},
Yinghan Hou\IEEEauthorrefmark{2}\IEEEauthorrefmark{3},
Botao Su\IEEEauthorrefmark{3}, and
Ercheng Zhao\IEEEauthorrefmark{3}}
\IEEEauthorblockA{\IEEEauthorrefmark{1}Dyson School of Design Engineering, Imperial College London, London, United Kingdom}
\IEEEauthorblockA{\IEEEauthorrefmark{2}Department of Electrical and Electronic Engineering, Imperial College London, London, United Kingdom}
\IEEEauthorblockA{\IEEEauthorrefmark{3}NOVASTACK PTE. LTD., Singapore\\
zy2926@ic.ac.uk, yh24@ic.ac.uk, botao.su@thundersoft.com, ercheng.zhao@thundersoft.com}}

\begin{document}
\maketitle

\begin{abstract}
Mixture-of-Agents systems combine several language models into a stronger response, but tool-using agents act over long horizons in which every step changes external state and the history keeps growing. We present Fusion-MoA, a serving architecture that brings multi-model collaboration into this loop through executor-anchored consultation. One Executor model holds the full context and emits every text and tool call; when consultation is admitted, three specialist models review its proposed next action through a compact view under a shared 120-second deadline. The fleet sits behind one OpenAI-compatible endpoint, so an unmodified agent client drives it as a single model. Fusion-MoA serves an in-car assistant and an in-car personal agent for intelligent cockpits and is deployed on MetaX C550 servers. Evaluated as four quantized open-weight models on five NVIDIA L40S GPUs and driven by Claude Code, it passes 78 of the 89 Terminal-Bench 2.1 tasks, 16 more than its Executor alone and 29 more than a self-hosted quantized Kimi K3 on the same client. On the 63 tasks that both it and the API-served LongCat-2.5-Preview pass, it needs 14.5 minutes on average against 17.5. Specialist reviews use 0.4\% of the Executor's tokens, and every tool call comes from the Executor.
\end{abstract}

\section{Introduction}
A terminal agent that compiles a legacy program, patches a vulnerable web
framework, or trains a small model works through a long loop of actions.
Each tool call installs a package, edits a file, or starts a process, and
each observation extends the history that grounds the next call. In our
Terminal-Bench runs a single task spans up to 140 executed tool calls and 5.0
hours, and the acting model re-reads 61k prompt tokens per call on average.

Such loops are reaching production assistants. Fusion-MoA is built for the
intelligent-cockpit services of \ifanon an automotive software provider\else
ThunderSoft\fi: an in-car assistant that answers questions and looks up travel
and schedule information, and an in-car personal agent that plans multi-step
trips, processes information across services, and runs tool-execution
workflows. Automakers require source code and user data to stay within their
compliance boundary, so both services share one locally deployed inference
backend on MetaX C550 servers instead of external APIs. The personal agent is
a long-horizon tool loop of the kind described above.

Mixture-of-Agents (MoA) combines several language models into a response
stronger than any single proposer \citep{moa2025}, and ranking-and-fusion
ensembles, debate, and role-based multi-agent frameworks extend the idea
\citep{llmblender2023,debate2023,camel2023,autogen2024,magenticone2024}. These
methods aggregate or exchange complete responses. A long-horizon tool loop
adds three requirements. Tool calls change external state, so once one
proposed call runs, the step needs a single action writer. A useful review
depends on the current execution state, and copying the growing history to
every collaborator multiplies the dominant inference cost at every step.
Consultation also sits on the critical path of a loop with a wall-clock budget
and an established client protocol, so the loop has to keep moving when a
collaborator is slow, and the client should keep calling one model.

\begin{figure*}[t]
\centering
\includegraphics[width=\textwidth]{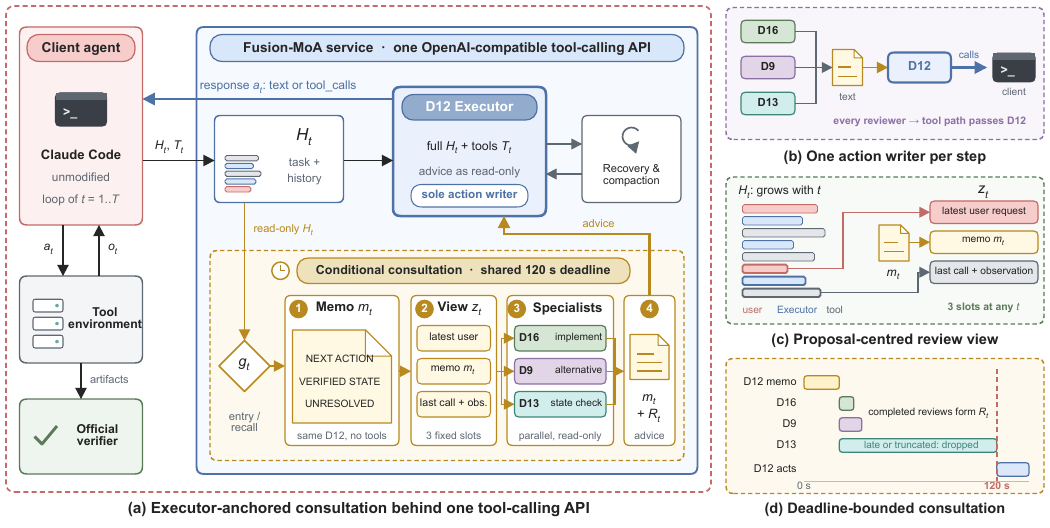}
\caption{Fusion-MoA. (a) An unmodified client sends the history $H_t$ and
tool schema $T_t$ to one tool-calling API. \Exec{}, the Executor, holds the full
context and returns every response $a_t$; the client executes tool calls and
appends observations $o_t$, and the official verifier scores the final task
artifacts. When the admission predicate $g_t$ fires, \Exec{} writes a memo
$m_t$~\circled{1}, the service builds the review view $z_t$~\circled{2},
three specialists review it in parallel~\circled{3}, and the memo and
completed reviews $R_t$ return to \Exec{} as read-only advice~\circled{4}.
(b) Specialist output reaches the client only through a new \Exec{} generation.
(c) The review view keeps three slots while $H_t$ grows. (d) Memo and reviews
share one 120-second deadline; late or truncated reviews are dropped.}
\label{fig:architecture}
\end{figure*}

Fusion-MoA meets these requirements with \emph{executor-anchored
consultation} (Figure~\ref{fig:architecture}). One Executor model holds the
full effective context and is the only component that emits text or tool
calls. When consultation is admitted, at task entry or on a recall event, the
Executor writes a three-part action memo; three specialist models review it
through a view that contains the latest user request, the memo, and the last
tool exchange, with no copy of the history; and the Executor generates the
next action from its full context, the memo, and the reviews that complete
within a shared 120-second deadline. The fleet is served behind one
OpenAI-compatible tool-calling endpoint, and Claude Code drives it without
modification.

We deploy four quantized open-weight models on five L40S GPUs and evaluate them on
the complete 89-task Terminal-Bench~2.1 inventory
\citep{terminalbench2026,terminalbench21release}, reporting each task's
official outcome together with the Agent time spent on it. Fusion-MoA passes 78
tasks at a median of 5.9 Agent minutes per solved task
(Figure~\ref{fig:results}a). On the same client, its Executor alone passes 62
tasks at 3.7 minutes and a self-hosted quantized Kimi K3 passes 49 at 4.3; the
API-served LongCat-2.5-Preview passes 69 at 10.8 minutes, and Terminus-2 runs of
DeepSeek V4 Flash, HY3, and LongCat-2.0 pass 45 to 61 tasks at 5.3 to 10.6
minutes. On the tasks it shares with each reference, Fusion-MoA spends a few
more minutes than the fast same-client runs and less time on average than
LongCat-2.5-Preview and the DeepSeek, HY3, and LongCat-2.0 runs
(Figure~\ref{fig:results}b). Traces
show the cost and reach of consultation: a specialist call reads about 1.2k prompt
tokens against 61k for an Executor call, 191 of 222 rounds finish before the
deadline, and departing advice leaves the Executor's plan adherence at its
usual level.

Our contributions are:
(1) \textbf{authority-bounded collective reasoning} through executor-anchored
consultation: several models reason about each step, one Executor writes every
action, and review runs on a compact view under conditional admission and a
shared deadline;
(2) \textbf{modular, memory-efficient heterogeneous serving}: independently
deployed, quantized Cells behind one OpenAI-compatible API, so domain
specialists plug in by configuration, the fleet fits on 48~GB GPUs, and an
unmodified agent client drives it; and
(3) \textbf{a full-inventory evaluation} that places accuracy next to Agent
time against same-client and cross-agent references, with a trace-level
account of how specialist advice reaches the action stream.

\section{Related Work}
\paragraph{Tool-using agents.} ReAct interleaves reasoning with actions and
observations \citep{react2022}; Toolformer and ToolLLM train models to call tools
and real-world APIs \citep{toolformer2023,toolllm2024}, and CodeAct uses
executable code as the action space \citep{codeact2024}; Reflexion and
Self-Refine add feedback-driven revision \citep{reflexion2023,selfrefine2023};
SWE-agent and OpenHands build agent--computer interfaces and open platforms for
software tasks \citep{sweagent2024,openhands2025}. Each runs
one model inside the loop. Fusion-MoA keeps the loop and its client intact and
places collaboration inside the model endpoint.

\paragraph{Multi-model collaboration.} Self-consistency and Tree of Thoughts
sample or search over one model's reasoning paths
\citep{selfconsistency2023,tot2023}. MoA layers aggregate proposer
responses \citep{moa2025}, and model composition shapes its gains
\citep{rethinkingmoa2025}. LLM-Blender ranks and fuses candidates
\citep{llmblender2023}; debate revises answers across rounds
\citep{debate2023}; routing and cascades pick one model per query for cost and
quality \citep{frugalgpt2023,routellm2024}. CAMEL, AutoGen, and Magentic-One
coordinate several acting agents, and MetaGPT and ChatDev assign software roles
to communicating agents
\citep{camel2023,autogen2024,magenticone2024,metagpt2024,chatdev2024}.
Fusion-MoA applies collaboration to the next action of a running tool loop:
specialists review it read-only, and one Executor emits it.

\paragraph{Context and serving.} MemGPT pages context between memory tiers
\citep{memgpt2023}, models use long contexts unevenly \citep{lostmiddle2024},
and compressing an agent's control context can break tool execution
\citep{controlcompression2026}; Fusion-MoA keeps the full history with the
Executor and gives each collaborator a step-local view. Continuous batching,
paged attention, structured-program execution, and speculative decoding
accelerate individual model servers
\citep{orca2022,vllm2023,sglang2024,speculativedecoding2023}; Fusion-MoA
composes such endpoints into a task-level service. Agent benchmarks score the
combined model, client, and environment
\citep{agentbench2023,webarena2024,swebench2023,terminalbench2026}.

\section{Executor-Anchored Consultation}
\subsection{Tool Loop and Fleet}
A client agent runs a loop $t=1,\dots,T$. At step $t$ it sends the endpoint
the message history $h_t$ and tool schema $T_t$, receives a response $a_t$
with text or tool calls, executes the calls, and appends the observations
$o_t$. The task ends when the client stops or its Agent time limit expires;
an official verifier then scores the artifacts.

Behind the endpoint, a \emph{Fleet} of independently served model endpoints,
called \emph{Cells}, carries out the request, and a \emph{Profile} assigns
roles and the consultation policy. The deployed Profile uses four Cells on five
NVIDIA L40S GPUs (Table~\ref{tab:cells}): \Exec{} is the Executor, and D16, D9,
and D13 review its proposals. A protocol bridge preserves streaming
termination and tool-call envelopes for the Claude Code client.

\begin{table}[t]
\centering\small\setlength{\tabcolsep}{3pt}
\begin{tabularx}{\columnwidth}{@{}lXl@{}}
\toprule
Cell & Model; quantization; GPUs & Role\\
\midrule
D12 & Qwen3.8-Flash-Next; IQ3\_S; 2 & Executor\\
D16 & Qwopus3.8-27B-Flash-V2; Q8\_0; 1 & Implementation\\
D9 & Qwopus3.6-27B-v2; Q6\_K; 1 & Alternatives\\
D13 & Muse-Glimmer-30B; Q4\_K\_XL; 1 & State check\\
\bottomrule
\end{tabularx}
\caption{Cells of the deployed Profile.}
\label{tab:cells}
\end{table}

\subsection{One Action Writer}
Let $H_t$ denote the Executor's \emph{effective} context: the task and its
retained history. \Exec{} receives $H_t$ and the client's tools $T_t$ on every
step. Specialists receive no tool bindings, and their text enters \Exec{} only
through a read-only advice channel. With $e$ the Executor and $\mathcal A$ the
specialists, the request path satisfies
\begin{equation}
\begin{split}
\left|\{c:\operatorname{emitsAction}(c)\}\right|&=1,\\
\forall r\in\mathcal A,\ \forall\rho:r\to^{*}\mathrm{Tools}:\ e&\in\rho .
\end{split}
\label{eq:singlewriter}
\end{equation}
Every public response is a normal \Exec{} generation returned as text or native
\texttt{tool\_calls}. Appendix~\ref{app:runtime} shows that this holds on
recovery, compaction, and failure paths, including when a specialist emits
tool-shaped text.

\subsection{Proposal-Centred Review}
When consultation is admitted, \Exec{} first writes a memo $m_t$ from a
read-only rendering of $H_t$ without tool bindings. The memo has three fields:
\texttt{NEXT ACTION}, \texttt{VERIFIED STATE} (observed facts supporting that
action), and \texttt{UNRESOLVED}. The service then builds the review view
\begin{equation}
z_t=\bigl(u(h_t),\;m_t,\;x(H_t)\bigr),
\label{eq:reviewview}
\end{equation}
where $u(h_t)$ is the latest user request and $x(H_t)$ the last tool
exchange, which pairs the latest observation with its originating call. The
view has three slots at every step and no copy of the history, so a
specialist's input depends on the current step only.

Each specialist receives the same $z_t$ independently. D16 proposes the
smallest implementation change to \texttt{NEXT ACTION}; D9 offers a materially
different path when the verified state supports one; D13 checks the action
against the requirements and the latest observation. Each role has an exact
no-op reply (\texttt{NO\_CHANGE}, \texttt{NO\_ALTERNATIVE}, \texttt{STATE\_OK}).
With $R_t$ the complete reviews available at the deadline, \Exec{} generates
\begin{equation}
a_t=\operatorname{D12}\bigl(H_t,\ \operatorname{advice}(m_t,R_t),\ T_t\bigr).
\label{eq:executor}
\end{equation}
The advice frames the memo as a revisable proposal and the reviews as
suggestions; \Exec{} may adopt part of them, set them aside, or continue.

\subsection{Bounded, Conditional Consultation}
A configured predicate $g_t$ admits consultation at task entry and on recall
events: a periodic check after every 16 tool cycles and detected repetition of
actions or results. It reads message structure and task identity and contains
no benchmark task names. An admitted step issues $k{+}2$ model calls for $k$
specialists (memo, reviews, final Executor call); any other step issues one.
Memo and reviews share one absolute 120-second allowance; reviews that finish
in time form $R_t$, and late or truncated ones are dropped. A failed memo
returns the step to direct execution. Bounded recovery paths regenerate after
a hard stall, continue a length-stopped response once, or summarize
intermediate history on a context-length error (Appendix~\ref{app:runtime});
each returns the public response to \Exec{}.

\section{Deployment and Evaluation}
\paragraph{Benchmark and client.} We use the 89-task Terminal-Bench~2.1
inventory at dataset revision \texttt{5c8eadf1f393}, Claude Code 2.1.220,
Harbor 0.15.0, and LiteLLM 1.93.0. Each trial runs in an isolated environment
and is scored by the task's official verifier. Original Agent limits range
from 600 to 12,000 seconds (median 900). An official reward of 1 is PASS, 0
is FAIL, and no numeric reward is UNSCORED. Because execution states and
correctness answer different questions \citep{accuracyhides2026}, Agent
timeouts and harness exceptions are recorded separately and never overwrite a
reward.

\paragraph{Agent time.} We report each task's official outcome and the
Agent time of the run that produced it: the wall-clock duration of the agent
phase recorded by the harness, excluding environment setup and verification.
Verifier limits and per-request output caps stay fixed; Appendix~\ref{app:protocol}
gives the protocol of every configuration.

\paragraph{Deployment on MetaX.} Fusion-MoA runs on MetaX C550 servers, the
hardware of its production cluster. Earlier versions of the system completed
the full 89-task Terminal-Bench~2.1 inventory on an eight-GPU C550 server with
Terminus-2, OpenCode, and Claude Code clients (Appendix~\ref{app:earlier}). The
current Profile runs on one C550 server with the Executor in W8A8 on four GPUs
(262,144-token context) and the three specialists in BF16 on one GPU each
(131,072-token contexts); together with the Fusion-MoA gateway, protocol
bridge, Claude Code client, and official verifiers, it completes
Terminal-Bench tool-execution tasks end to end, with runs that pass the
official verifier. The production cluster has 64 such servers, each with eight C550 GPUs (64~GiB
each, 512~GiB per server), 160 logical CPUs, 1~TiB of memory, and two data disks
of about 7~TiB: 512 GPUs with 32~TiB of GPU
memory, 10,240 logical CPUs, 64~TiB of system memory, and about 896~TiB of raw
storage. Table~\ref{tab:workloads} maps the two cockpit workloads to
Fusion-MoA's paths. The full-inventory evaluation below runs the same four
models with GGUF quantization on five L40S GPUs (Table~\ref{tab:cells}), and
Terminal-Bench~2.1 stands in for the personal agent's long tool-execution
workflows.

\begin{table}[t]
\centering\small\setlength{\tabcolsep}{3pt}\renewcommand{\tabularxcolumn}[1]{>{\raggedright\arraybackslash}p{#1}}
\begin{tabularx}{\columnwidth}{@{}p{1.45cm}XX@{}}
\toprule
Workload & Typical requests & Fusion-MoA path\\
\midrule
In-car assistant & Question answering, travel and schedule lookups & One model API; fast Executor path\\
\addlinespace
In-car personal agent & Multi-step trip planning, cross-service information processing, tool-execution workflows & On-demand consultation; long-horizon state; single tool-action writer\\
\bottomrule
\end{tabularx}
\caption{Intelligent-cockpit workloads served by Fusion-MoA.}
\label{tab:workloads}
\end{table}

\paragraph{Configurations.} \emph{Fusion-MoA} runs the full Profile.
\emph{Executor-only} keeps the same \Exec{} checkpoint, managed execution, and
recovery, and disables the memo and the specialists; it runs once per task
within the original limits.

\paragraph{References.} \emph{Same client:} Kimi K3, a quantized deployment
on in-house NVIDIA B200 GPUs at maximum reasoning effort, runs once per task within the original
limits on the same runtime and task revision. LongCat-2.5-Preview runs through
its official API at maximum reasoning effort on the same runtime, task
revision, and environment images, once per task with an extended Agent
allowance and unchanged verifier limits. \emph{Terminus-2:} self-deployed DeepSeek V4
Flash 0731 and HY3, and LongCat-2.0 through its official API, ran with
Terminus~2.0.0, once per task, a 10,800-second Agent allowance, temperature
0.7, and two concurrent tasks per model; HY3 also ran once with extended Agent
and verifier limits, and a self-deployed Qwen3.6-35B-A3B once within the original limits with
six concurrent tasks. Appendix~\ref{app:protocol} lists the protocol details.

\section{Results}
\subsection{Accuracy and Agent Time}
\begin{figure*}[t]
\centering
\includegraphics[width=\textwidth]{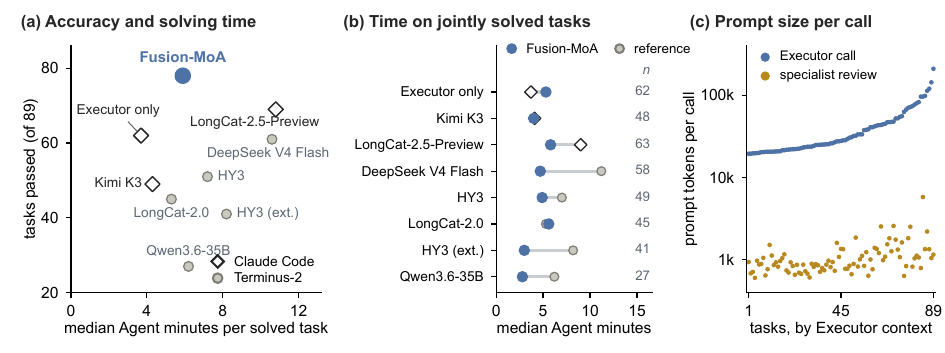}
\caption{Accuracy and time on Terminal-Bench~2.1. (a) Tasks passed against
median Agent minutes per solved task. (b) Median Agent minutes on the tasks
that Fusion-MoA and each reference both pass; $n$ is the number of such tasks.
(c) Mean prompt tokens per Executor call and per specialist review for each of
the 89 tasks, sorted by Executor context.}
\label{fig:results}
\end{figure*}

Fusion-MoA passes 78 of the 89 tasks, with every task scored
(Table~\ref{tab:coverage}). Figure~\ref{fig:results}a places accuracy against
the median Agent time per solved task. The two fast same-client runs solve a
task in about four minutes at the median (means 8.4 and 7.3), with 62 tasks for
Fusion-MoA's Executor alone and 49 for Kimi K3; Fusion-MoA takes 5.9 minutes
(mean 15.2) and passes 78. With this time,
Fusion-MoA solves 16 tasks its Executor alone misses and 30 that Kimi K3
misses, and each of them solves at most one task outside Fusion-MoA's set.
LongCat-2.5-Preview passes 69 tasks at 10.8 minutes (mean 21.9), and the
Terminus-2 references pass 27 to 61 tasks at medians of 5.3 to 10.6 minutes
(means 10.2 to 27.5).

Solving time per task shows where the time goes (Figure~\ref{fig:results}b).
On the tasks both solve, Fusion-MoA's median is 5.3 minutes against 3.7 for
its Executor alone (means 12.5 and 8.4) and 4.0 against 4.1 for Kimi K3 (means
10.8 and 6.8), so Fusion-MoA spends a few more minutes on tasks that a fast run
also solves. Against LongCat-2.5-Preview and the Terminus-2 runs of DeepSeek V4
Flash, HY3, and LongCat-2.0, it is faster on shared tasks on average: 14.5 against 17.5 minutes for LongCat-2.5-Preview, 13.6 against
22.8 for DeepSeek V4 Flash, 13.9 against 18.0 for HY3, and 14.2 against 27.5
for LongCat-2.0.

\begin{table}[t]
\centering\small\setlength{\tabcolsep}{3.2pt}
\begin{tabularx}{\columnwidth}{@{}Xrrrr@{}}
\toprule
Run & PASS & U & Median & Mean\\
\midrule
\multicolumn{5}{@{}l}{\emph{Claude Code}}\\
Fusion-MoA & \textbf{78} & 0 & 5.9 & 15.2\\
Executor only (\Exec{}) & 62 & 2 & 3.7 & 8.4\\
Kimi K3 (self-hosted) & 49 & 1 & 4.3 & 7.3\\
LongCat-2.5-Preview & 69 & 0 & 10.8 & 21.9\\
\midrule
\multicolumn{5}{@{}l}{\emph{Terminus-2}}\\
DeepSeek V4 Flash 0731 & 61 & 1 & 10.6 & 22.3\\
HY3 & 51 & 2 & 7.2 & 18.2\\
LongCat-2.0 & 45 & 2 & 5.3 & 27.5\\
HY3, extended limits & 41 & 2 & 8.2 & 16.8\\
Qwen3.6-35B-A3B & 27 & 6 & 6.2 & 10.2\\
\bottomrule
\end{tabularx}
\caption{Official outcomes and Agent time on the 89 tasks. U: unscored.
Median, Mean: Agent minutes per passed task. With the runtime assets of six
tasks restored, DeepSeek, HY3, and LongCat-2.0 pass 62, 56, and 46
(Appendix~\ref{app:protocol}).}
\label{tab:coverage}
\end{table}

\subsection{The Consultation Path in Operation}
This section and the next analyse the gateway and wire logs of the 89
Fusion-MoA runs behind Table~\ref{tab:coverage} (Table~\ref{tab:consult}).
Of the 222 rounds, 89 open a task and 133 are recalls; \Exec{} delivered the
memo in 186, the median round took 43.3 seconds, and 191 finished before the
deadline. Review stays
small as context grows: a mean Executor call reads 19k to 211k prompt tokens
per task, while a specialist call reads 0.6k to 5.8k
(Figure~\ref{fig:results}c). Specialist calls are 20\% of \Exec{}'s calls and
use 0.4\% of its tokens. Every response carrying tool calls came from \Exec{}
(2,482 responses); none of the 561 specialist requests offered tools, and all
2,632 executed tool calls match the ID of a \Exec{} response.

\subsection{What Specialist Advice Changes}
\label{sec:advice}
Advice reaches the action stream through one request, the Executor call that
follows the reviews. We read every consultation round whose reviews contain
substantive advice, with the response that follows it, and compare them with
recall rounds whose reviews only confirm the plan. Because an LLM judgment can
move with the judge \citep{llmjudge2023,judgechanges2026}, two independent
annotation passes labelled whether the advice departs from the memo's
\texttt{NEXT ACTION} (18 of 20 agree, Cohen's $\kappa=0.78$;
\citealp{cohen1960}) and what the response did with
it (15 of 20, $\kappa=0.59$), and, blind to the reviews, whether each response
carries out \texttt{NEXT ACTION} (85 of 96, $\kappa=0.80$;
Appendix~\ref{app:advice}).

Of the 186 reviewed rounds, 20 carry substantive advice (23 of the 472 usable
reviews), all of them recalls deep in long runs. In eight the advice restates
\Exec{}'s plan; in 12 it proposes a different step, and \Exec{}'s next
action takes up two of these in part and keeps its plan in the other ten.
Departing advice leaves plan adherence at its level without advice: blind to
the reviews, the next action modifies or replaces \texttt{NEXT ACTION} in 3 of
the 12 departing rounds and in 30 of the 75 confirming recall rounds (one-sided
Fisher exact test, $p=0.91$; \citealp{fisher1922}). A specialist can thus add
information to a running task while the Executor keeps a coherent plan. The
advice \Exec{} took up carries specific knowledge. In \path{protein-assembly}, after a
GraphQL lookup failed with a \texttt{KeyError}, D16 proposed printing the raw
response body; \Exec{} did so, the body exposed a schema error, and within two
calls \Exec{} rewrote the query and retrieved the sequences the task needed.

\section{Lessons for Deployment}
\paragraph{Authority-bounded collective reasoning.} Routing every public
response through one Executor let an unmodified commercial client run a
heterogeneous fleet as a drop-in model endpoint while four models reason about
each consulted step.

\paragraph{Plug-and-play domain specialists.} Specialists are bound only by
the review view and their no-op reply, so swapping one is a Profile change: the
deployed fleet already mixes three model lines and four quantization formats
(Table~\ref{tab:cells}), and earlier versions ran the same roles with different
Executors and clients (Appendix~\ref{app:earlier}). A cockpit deployment can
add domain-tuned specialists, such as a vehicle-domain or tool-use model,
without retraining the Executor or changing the agent's tool interface
(Appendix~\ref{app:specialists}).

\paragraph{Memory-efficient heterogeneous serving.} Independent, quantized
Cells keep each specialist on one 48~GB L40S GPU and the Executor on two, so
collaboration needs no merged checkpoint.

\paragraph{Review the proposal under a deadline.} Grounding review in the
Executor's memo kept a specialist call near 1.2k tokens while Executor contexts
grew past 200k, and a shared deadline that drops late reviews kept 191 of 222
rounds within 120 seconds.

\paragraph{Match the path to the workload.} The fast Executor-only path (62
tasks, median 3.7 minutes per solved task) suits interactive assistant
requests, and the full system (78 tasks, 5.9 minutes) suits personal-agent
workflows; the Profile selects the path per service behind the same API.

\section{Conclusion}
Fusion-MoA separates collective reasoning from execution authority:
specialists review a compact view of the Executor's proposal under a shared
deadline, and the Executor alone writes the action behind one tool-calling API
for in-car assistant and personal-agent services. With four quantized
open-weight models, it passes 78 of 89 Terminal-Bench~2.1 tasks and solves
tasks that fast single-model runs leave open.

\section*{Limitations}
The evaluation covers one deployment on one 89-task inventory with one record
per configuration, so it reports recorded coverage rather than
run-to-run success probabilities. The comparisons span different clients,
serving stacks, Agent allowances, concurrency, and dates; we place every run on
the common axis of Agent time, which also includes model serving latency.
Serving hardware differs as well: Fusion-MoA ran on five L40S GPUs, whereas
Kimi K3 ran on in-house B200 GPUs, and the other references on their own
self-deployed or provider infrastructure, so Agent time reflects hardware speed as well as the agent. Because some
Fusion-MoA runs on the L40S GPUs reached the Agent limit before finishing,
tasks not resolved within the original limits were run again with a longer
allowance; the Executor-only and Kimi K3 runs use the original limits. The task revision of the Terminus references was not independently established.
The advice analysis is observational and relies on AI-assisted annotation;
same-state replays that remove the advice would measure its causal effect. The
single-writer property concerns the origin of actions; their correctness and
safety depend on their content and on the client's tool runtime.
Although Fusion-MoA supports independent Cell replacement through Profile
updates, specialist selection is currently preconfigured rather than
dynamically adapted to each task. The system does not yet automatically select
and hot-swap domain-specialized models during execution, nor does it
autonomously evolve its model fleet from execution feedback. Task-adaptive
specialist selection and closed-loop fleet evolution are natural directions
for future work.

\section*{Ethical Considerations}
Fusion-MoA executes shell commands through its client. All runs used isolated
benchmark containers scored by official verifiers, and deployments outside
such sandboxes need execution controls and human oversight. The study involves
no human subjects or personal data; API models were used under their
providers' terms. AI-assisted tools were used for manuscript
editing, figure preparation, consistency checks, and the annotation passes in
Section~\ref{sec:advice}; the authors checked every reported number against
the run records and take responsibility for the content.

\bibliography{references}

\appendix

\section{Runtime Contract}
\label{app:runtime}
\paragraph{State and admission.} Write the request state as
$\sigma_t=(q,H_t,\iota_t,m_t,R_t,a_t)$, with runtime phase $q$, effective
history $H_t$, process-local task identity and admission state $\iota_t$,
optional memo $m_t$, available reviews $R_t$, and the \Exec{} output envelope
$a_t$. Admission is a configured predicate of initial-task eligibility,
stagnation state, and explicit recall; the initial claim is taken before
awaiting optional model calls, so simultaneous requests cannot claim the same
initial consultation. The normal phase paths are
$\mathrm{validate}\to\mathrm{execute}\to\mathrm{return}$ and
$\mathrm{validate}\to\mathrm{memo}\to\mathrm{review}\to\mathrm{execute}\to\mathrm{return}$.
A memo failure removes optional guidance; missing reviews may leave a valid
memo for the same Executor.

\paragraph{Sole action owner.} \emph{Proposition.} If the public response
boundary returns only the normal Executor output, reviewer outputs are rendered
only as read-only text, and the client executes only tool intent received
through that boundary, then every model-originated actionable output has
\Exec{} as its immediate model owner, including when optional reviewers fail.
\emph{Argument.} The only model-to-public-response edge is \Exec{}'s execution
edge. Reviewers have no tool binding or client edge; even a tool-shaped review
is serialized into advisory text, so every path from it to a client tool
traverses a new \Exec{} generation. Removing missing reviews cannot add an
edge. After a memo or review failure, the following public call remains
\Exec{}'s; if memo generation fails, guidance is empty and \Exec{} executes
directly.

\paragraph{Context and recovery.} On a context-length error, auxiliary
compaction keeps the task prefix and recent messages and summarizes the
intermediate history in chronological segments, one per specialist; the
checkpoint is task-local and allows at most two context-error retries. No
compaction event occurred in the evaluated runs. If optional guidance causes a
request-format error (400/422), the runtime removes it and retries the
Executor context. Clean Recovery regenerates from the task and selected cycle
evidence after a hard stall; Output Recovery continues a length-stopped
response once, up to 512 tokens. Executor-only execution keeps managed
execution and recovery and disables the memo, the specialists, and compaction.

\paragraph{Decoding.} \Exec{} uses a 262,144-token context, thinking at medium
effort, temperature 0.6, top-$p$ 0.95, top-$k$ 20, and a 16,384-token output
cap. Reviewers use 131,072-token single-slot endpoints, temperature 0.3, and
output caps of 768, 512, and 512 tokens; the memo is non-thinking with a
768-token cap.

\section{Protocol and Scoring}
\label{app:protocol}

\begin{table*}[t]
\centering\small\renewcommand{\tabularxcolumn}[1]{>{\raggedright\arraybackslash}p{#1}}
\begin{tabularx}{\textwidth}{@{}lXXX@{}}
\toprule
Property & Fusion-MoA & Claude Code references & Terminus references\\
\midrule
Agent & Claude Code 2.1.220 & Claude Code 2.1.220 & Terminus 2.0.0\\
Agent allowance & Original task limits; given the speed of the local hardware, tasks not resolved within them were run again with a longer allowance & Kimi K3: original limits; LongCat-2.5-Preview: extended & 10,800 s per task; HY3 also extended; Qwen3.6 original limits\\
Concurrency & One task & One task per model & Two tasks per model (Qwen3.6: six)\\
Sampling & \Exec{} $T=0.6$, $p=0.95$ & Provider default; maximum effort & $T=0.7$, $p=1.0$\\
Request output cap & \Exec{} 16,384 tokens & Client default & 8,192 tokens\\
Model access & Local GPUs & Kimi K3: quantized deployment on in-house B200 GPUs; LongCat-2.5-Preview: official API; both through a relay & DeepSeek, HY3, and Qwen3.6: self-deployed; LongCat-2.0: official API\\
Task revision & Pinned & Same pinned revision & Matching task IDs\\
Scoring & Official rewards & Official rewards & Official rewards\\
\bottomrule
\end{tabularx}
\caption{Protocol details for the runs in Table~\ref{tab:coverage}.}
\label{tab:protocols}
\end{table*}

\paragraph{Outcomes.} For an official record $r$, $s(r)$ is PASS if
$\mathrm{reward}(r)=1$, FAIL if $\mathrm{reward}(r)=0$, and UNSCORED if no
numeric reward exists. Agent and harness exceptions are a separate diagnostic
(Table~\ref{tab:exceptions}); an Agent timeout can coexist with an official
PASS. The HY3 extended-limit run recorded two
environment-start failures (\path{qemu-alpine-ssh}, \path{qemu-startup})
without a numeric reward; they are UNSCORED here.

\paragraph{Reference runs.} Six tasks lacked runtime assets in the
10,800-second Terminus runs (\path{build-pov-ray}, \path{hf-model-inference},
\path{mteb-leaderboard}, \path{mteb-retrieve}, \path{reshard-c4-data},
\path{protein-assembly}); with the assets restored, DeepSeek V4 Flash, HY3, and
LongCat-2.0 pass 5, 5, and 2 of them (4, 0, and 1 without the assets). Kimi K3's requests passed through an
authenticated relay, and a client fix that terminates the agent process in
cancelled containers took effect from its 30th task; the task revision,
limits, and verifier were unchanged. LongCat-2.5-Preview ran in October 2026
with one task at a time; its requests went to the official
Anthropic-compatible endpoint through a local relay. Qwen3.6-35B-A3B ran with
thinking disabled on a local Docker host; its five-trial mean is 30.1\%.

\paragraph{Shared tasks.} For Fusion-MoA and each reference, the tasks
passed by only one side are 16 and 0 (Executor only), 30 and 1 (Kimi K3), 15
and 6 (LongCat-2.5-Preview), 20 and 3 (DeepSeek V4 Flash), 29 and 2 (HY3), 33
and 0 (LongCat-2.0), 37 and 0 (HY3, extended limits), and 51 and 0
(Qwen3.6-35B-A3B). Figure~\ref{fig:results}b uses the tasks passed by both.

\begin{table}[t]
\centering\small\setlength{\tabcolsep}{3pt}
\begin{tabularx}{\columnwidth}{@{}Xrr@{}}
\toprule
Run & Exceptions & PASS with one\\
\midrule
Fusion-MoA & 13 & 7\\
Executor only & 20 & 4\\
Kimi K3 & 15 & 1\\
LongCat-2.5-Preview & 11 & 4\\
DeepSeek V4 Flash (Terminus) & 8 & 0\\
HY3 (Terminus) & 11 & 0\\
LongCat-2.0 (Terminus) & 11 & 1\\
HY3, extended limits & 19 & 0\\
Qwen3.6-35B-A3B & 50 & 3\\
\bottomrule
\end{tabularx}
\caption{Exception diagnostics overlap official outcomes; they are not extra
outcome categories.}
\label{tab:exceptions}
\end{table}

\section{Consultation Statistics}
\label{app:consult}

\begin{table}[t]
\centering\small
\begin{tabularx}{\columnwidth}{@{}Xr@{}}
\toprule
Measure & Value\\
\midrule
Rounds: task entry / recall & 89 / 133\\
Tasks with a recall round & 51 of 89\\
Memo delivered & 186 of 222\\
Rounds with $\geq$2 usable reviews & 181 of 186\\
Round time, median / p90 & 43.3 / 120.0 s\\
Rounds completed before 120 s & 191 of 222\\
Prompt per call: Executor / memo / review & 61.1k / 45.3k / 1.2k\\
Specialist share of \Exec{} tokens & 0.4\%\\
Usable reviews: no-op / substantive & 449 / 23\\
Rounds with substantive advice (all recalls) & 20\\
Specialist requests / with tools offered & 561 / 0\\
Tool-call responses from \Exec{} & 2,482 of 2,482\\
Executed tool calls matched to \Exec{} & 2,632 of 2,632\\
\bottomrule
\end{tabularx}
\caption{Consultation path in the Fusion-MoA runs behind the outcomes in
Table~\ref{tab:coverage} (89 runs). A review is no-op when it is exactly the role's no-op
reply; every other usable review is substantive. Three further rounds were in
flight when the Agent limit ended their session, and specialist requests
include those rounds. In 36 rounds the memo fell back to direct execution (27
at the deadline, 9 at the context limit). Of the 133 recalls, 112 came from
the periodic check.}
\label{tab:consult}
\end{table}

The memo accounts for about four fifths of round time; the slowest
specialist in a round took a median of 13.1 seconds. Memo and reviews together
use 5.5\% of all tokens.

\section{Advice Annotation}
\label{app:advice}
These 89 runs contain 20 delivered rounds with at least one substantive
review, 23 reviews in all (Table~\ref{tab:advice}). For each, the annotation packet
holds the memo, the substantive reviews, the guided Executor response (text,
tool calls, and results), and the next five Executor steps; task outcomes are
omitted. Two independent AI-assisted passes used the same written rubric.
\emph{Relation:} the review departs from the memo's \texttt{NEXT ACTION} if it
proposes a different action, an added check, another tool or approach, or a
factual correction, and matches it otherwise. \emph{Response:} the guided
response adopts the departing part, adopts part of it, or keeps \Exec{}'s own
plan; for matching advice it follows the shared plan or takes another step.
Disagreements resolve to the weaker label.

The plan-adherence comparison mixes the 20 rounds with the 76 recall rounds
whose delivered reviews were all no-op replies; one of these has an empty
response and is left out.
Each item shows only the memo and the next response, with reviews removed and
order shuffled; two further passes labelled whether the response follows,
partly follows, or departs from \texttt{NEXT ACTION}, and disagreements resolve
toward adherence.

\begin{table}[t]
\centering\scriptsize\setlength{\tabcolsep}{2.8pt}
\begin{tabularx}{\columnwidth}{@{}lXllr@{}}
\toprule
& Task & Review & Advice & Response\\
\midrule
R00 & \path{protein-assembly} & D16, D9 & departs & partly\\
R01 & \path{build-cython-ext} & D9 & departs & kept plan\\
R02 & \path{build-cython-ext} & D9 & matches & same plan\\
R03 & \path{crack-7z-hash} & D9 & matches & other step\\
R04 & \path{git-multibranch} & D9 & departs & kept plan\\
R05 & \path{mailman} & D9 & departs & kept plan\\
R06 & \path{mailman} & D9 & departs & kept plan\\
R07 & \path{adaptive-rejection-sampler} & D16 & matches & same plan\\
R08 & \path{caffe-cifar-10} & D9 & matches & same plan\\
R09 & \path{caffe-cifar-10} & D9 & departs & kept plan\\
R10 & \path{caffe-cifar-10} & D9 & departs & kept plan\\
R11 & \path{compile-compcert} & D9 & matches & same plan\\
R12 & \path{filter-js-from-html} & D16 & matches & same plan\\
R13 & \path{gcode-to-text} & D9 & matches & other step\\
R14 & \path{gpt2-codegolf} & D16, D9 & departs & partly\\
R15 & \path{install-windows-3.11} & D16, D9 & departs & kept plan\\
R16 & \path{install-windows-3.11} & D9 & matches & same plan\\
R17 & \path{make-doom-for-mips} & D9 & departs & kept plan\\
R18 & \path{train-fasttext} & D16 & departs & kept plan\\
R19 & \path{winning-avg-corewars} & D9 & departs & kept plan\\
\bottomrule
\end{tabularx}
\caption{Consultation rounds with substantive advice in the 89 runs, in log
order.}
\label{tab:advice}
\end{table}

\section{Plug-and-Play Domain Specialists}
\label{app:specialists}
Each specialist is a Cell with its own endpoint, checkpoint, and quantization,
and the Profile binds it to a review role. Adding or replacing a specialist
therefore changes the Profile and leaves the Executor, the client, and the
tool schema unchanged; Cells are replaced independently behind the stable API.
Table~\ref{tab:specialists} lists example roles for an intelligent cockpit.
These configurations illustrate the extension path and are not evaluated in
this paper.

\begin{table}[t]
\centering\small\setlength{\tabcolsep}{3pt}\renewcommand{\tabularxcolumn}[1]{>{\raggedright\arraybackslash}p{#1}}
\begin{tabularx}{\columnwidth}{@{}p{1.5cm}XX@{}}
\toprule
Specialist & Candidate model & Role in review\\
\midrule
General reasoning & General open-weight LLM & Reasoning on complex tasks\\
\addlinespace
Vehicle domain & Automotive-domain fine-tuned model & Vehicle knowledge, control intents, feature constraints\\
\addlinespace
Trip planning & Planning-tuned model & Multi-step itinerary planning\\
\addlinespace
Tool use & Tool-use fine-tuned model & API argument checks, tool-execution suggestions\\
\bottomrule
\end{tabularx}
\caption{Example specialist roles for an intelligent cockpit (illustrative
configurations, not evaluated here).}
\label{tab:specialists}
\end{table}

\section{Earlier Fusion-MoA Versions}
\label{app:earlier}
Earlier versions ran the same 89 tasks on an eight-GPU MetaX C550 server with other
clients, fixed Agent allowances, two concurrent tasks, and one run per task
(Table~\ref{tab:earlier}); infrastructure errors count as non-PASS.

\begin{table}[t]
\centering\scriptsize\setlength{\tabcolsep}{2.8pt}
\begin{tabularx}{\columnwidth}{@{}Xlrrr@{}}
\toprule
Executor and client & Allowance & PASS & FAIL & Infra\\
\midrule
Qwopus3.6-27B-Coder, Terminus~2 & 14,400 s & 45 & 27 & 17\\
Qwen3.8-27B, OpenCode (medium) & 18,000 s & 42 & 42 & 5\\
Qwen3.8-27B, OpenCode (xhigh) & 18,000 s & 31 & 49 & 9\\
Qwen3.8-27B, Claude Code (xhigh) & 18,000 s & 31 & 29 & 29\\
\bottomrule
\end{tabularx}
\caption{Earlier Fusion-MoA runs on the 89 tasks, one run per task. Infra:
unscored or infrastructure error.}
\label{tab:earlier}
\end{table}

\section{Artifact}
\label{app:artifact}
\ifanon
We will release the runtime, protocol bridge, offline tests with mocked
backends, and a per-task audit (an 89-task scalar ledger, official-record
scalars, and a standard-library audit script) upon publication. The release
excludes credentials, deployment addresses, model weights, benchmark task
bodies, and raw traces.
\else
The runtime, protocol bridge, offline tests with mocked backends, and a
per-task audit (an 89-task scalar ledger, official-record scalars, and a
standard-library audit script) are released at
\url{https://github.com/tongjiu123/Fusion-MoA-Pioneer} under
\texttt{versions/v0.9.4.8} (code revision \texttt{8273f3be08c2}). The release
excludes credentials, deployment addresses, model weights, benchmark task
bodies, and raw traces.
\fi

\end{document}